\documentstyle[axodraw,epic,aps]{revtex}
\begin{document}
\title{The velocities of light in modified QED 
vacua\thanks{Talk delivered at the workshop ``Superluminal(?) Velocities'',
Cologne, July 7-10, 1998}}
\author{K. Scharnhorst\thanks{email: 
scharnh@physik.hu-berlin.de}}
\address{Humboldt-Universit\"at zu Berlin, Institut f\"ur Physik,
Invalidenstr.\ 110, D-10115 Berlin,
Federal Republic of Germany}
\maketitle
\begin {abstract}
QED vacua under the influence of external conditions (background fields,
finite temperature, boundary conditions) can be
considered as dispersive media whose complex behaviour can no longer 
be described in terms of a single universal vacuum velocity of light
$c$. Beginning in the early 1950's (J.S.\ Toll), quantum field theoretic
investigations have led to considerable insight into the relation 
between the vacuum structure and the propagation of light. 
Recent years have witnessed a significant growth of activity
in this field of research.
After a short overview, two characteristic situations are discussed:
the propagation of light in a constant homogeneous 
magnetic field and in a Casimir
vacuum. The latter appears to be particularly interesting because the 
Casimir vacuum has been found to exhibit modes of the propagation of light
with phase and group velocities larger than $c$ in the low frequency 
domain $\omega\ll m$ where $m$ is the electron mass.
The impact of this result on the front velocity of light in a Casimir
vacuum is discussed by means of the Kramers-Kronig relation.
\end{abstract}
\section{Introduction}

Relativistic quantum field theory can be understood as having 
emerged, historically, from a combination of special relativity
and quantum mechanics, but as its very fathers 
have always been very aware it is 
not a synthesis of these two theories. In standard relativistic
quantum field theory, space-time is considered as a fixed arena
in which the physical processes take place.
The characteristics of the propagation of 
light are considered as {\it classical} input to the theory.
This entails the view that there exists a single universal
vacuum velocity of light $c$. However, starting in the early 
1950's \cite{toll} research in quantum electrodynamics (QED)
has revealed that higher conceptional sophistication in 
discussing the propagation of light in a vacuum is required.
This more advanced insight derives from studies of QED vacua
which have been modified by means of external conditions [background
fields (electromagnetic, gravitational), finite temperature
(heatbath), boundary conditions]. These modified vacua can 
be explored by studying the behaviour of particles 
immersed into the vacuum. One particular method, which is of 
special conceptional significance, consist in the investigation
of the propagation of photons (light) in the vacuum modified by
external conditions. The characteristics of the propagation of 
light then describe certain aspects of the vacuum structure.
As a result, it has been found that modified QED vacua are
complicated dispersive media which exhibit almost every phenomenon
which is well known from ordinary condensed matter media in this 
respect. This is a {\it quantum} effect which is caused by the 
phenomenon of vacuum polarization.
The propagation of light may depend on:
\begin{itemize}
\item[-]
the photon polarization,
\item[-]
the direction of propagation,
\item[-]
the photon frequency $\omega$ [(temporal) dispersion],
\item[-]
the space-time location.
\end{itemize}
The description of these dependencies 
can conveniently be performed by means of a refractive index
$n(\omega)$ for the appropriate modes of the propagation of 
light (we indicate the dependence on the photon frequency $\omega$ only;
unless specified otherwise, $n(\omega)$ denotes the real part 
of the refractive index throughout the paper).
Accordingly, as the refractive index of modified QED vacua is not a universal
constant one needs to consider different notions of light velocities which
have different physical significance (for a discussion of 
a number of notions of a propagation velocity see, e.g., \cite{bril,smit}).
In this article, we will concentrate our attention onto three
velocities, the phase velocity
\begin{equation}
v_{\rm ph}\ =\ \frac{c}{n(\omega)}\ \ ,
\end{equation}
the group velocity
\begin{equation}
v_{\rm gr}\ =\ \frac{c}{n_{gr}(\omega)}\ \ ,
\ \ n_{gr}(\omega)\ =\ n(\omega)\ +\ \omega 
\frac{\partial n(\omega)}{\partial\omega}\ \ ,
\end{equation}
and the front velocity
\begin{equation}
\label{A3}
v_{\rm fr}\ =\ \frac{c}{n(\infty)}\ \ ,
\end{equation}
which is the velocity related to the concept of the space-time
structure (note that, in general, in modified QED vacua $n(\infty)$ has to
be calculated and cannot be set equal to 1 just by default).

The behaviour of the different velocities can 
be discussed as soon as information about the refractive indices
for the different modes propagating in the modified QED vacua 
is available. The task of quantum field theory consists in the 
calculation of these refractive indices. To this end one has
to calculate the (renormalized) effective Maxwell action 
$\Gamma_{\rm eff}[{\bf E},{\bf B}]$ under
the external conditions under consideration
(${\bf E}$, ${\bf B}$ are the field strengths of the test wave 
which are assumed to be sufficiently small). Within quantum
field theory this is a well established procedure. The effective
Maxwell action provides us with a convenient interface between quantum
field theory and classical field theory. Once the 
effective Maxwell action is known, at least in some sensible
approximation, a dispersion analysis can be performed by
considering the effective equations of motion (effective 
Maxwell equations). This analysis proceeds along the same lines as 
the analysis for any other action would proceed, irrespective of its 
origin. 

Within QED (as in many quantum field theoretic models), the effective
(Maxwell) action can only explicitly be calculated, even at a given order
of perturbation theory, if certain approximations are applied. 
The most commonly
applied approximation is to consider the propagation of low 
frequency photons with $\omega\ll m$, i.e., of photons whose wavelength is 
much larger than the Compton wavelength of the electron ($m$ is the 
electron mass, $\hbar, c = 1$ throughout the paper unless inserted
explicitly). Below we list
the situations studied so far together with a qualitative overview
of the results obtained for the low frequency refractive index 
$n(0) = n\left(\omega\ (\ll m)\right)$. The references given
in the column `further literature' are not meant to represent a complete
list. This, in particular, concerns the investigation of the propagation
of light in electromagnetic background fields which has a rather
long and rich history.
$$
\begin{array}{lcll}
&\mbox{\small refractive index} &\mbox{\small first studied}&
\mbox{\small{further literature}}\\
-\ \mbox{background fields}& &&\\[2mm]
\ \ \ \ \ -\ \mbox{electromagnetic}   &n(0)\ge 1 & 
\mbox{1952 Toll \cite{toll}}&
\mbox{\cite{erbe1,erbe2,bial,adle,tsai1,tsai2,dani1,shor,ditt,gies}}
\\[3mm]
\ \ \ \ \ -\ \mbox{gravitational}   &n(0)\stackrel{\displaystyle\ge}{<} 1 & 
\mbox{1980 Drummond/Hathrell \cite{drum}}&
\mbox{\cite{dolg1,dani1,khri,dani2,shor,cai,dolg2,dolg3}}\\[1mm]
-\ \mbox{finite temperature}&n(0)> 1 &\mbox{1983 Tarrach \cite{tarr}, 
corrected: 1990 Barton \cite{bart1}}&
\mbox{\cite{bart2,lato,ditt,gies}}\\[1mm]
-\ \mbox{Casimir vacuum}&n(0)\le 1 &\mbox{1990 Scharnhorst \cite{scha}, 
Barton \cite{bart1}}&
\mbox{\cite{milo1,benm,bart3,lato,ditt,dolg3}}\\
\ \ \ \mbox{(boundary conditions)}&&&
\end{array}
$$
The above results have all been obtained by considering 1-loop or 2-loop 
contributions to the effective Maxwell action only (which in the 
low frequency approximation yield the leading contribution in many cases). 
Very few results are available, even in 1-loop or 2-loop 
approximation, for arbitrary photon frequencies $\omega$. However, as
one can see from Eq.\ (\ref{A3}), if one wants to discuss the 
front velocity of light, which is of greatest conceptional 
significance in discussing the space-time structure of the 
modified vacuum, one has to study the effective Maxwell action 
for arbitrarily large frequencies $\omega$ of the test photon exploring this
vacuum. As interesting as any 1-loop or 2-loop calculation for 
arbitrary frequencies $\omega$ may be, the calculation of $n(\infty)$ is a
truly nonperturbative task \cite{bart3}, p.\ 2041, 
which has not yet been solved in any of the situations listed above.

Latorre, Pascual and Tarrach \cite{lato} have proposed the following unified 
formula (sum rule) for the above listed results for the low 
frequency refractive index $n(0)$
which facilitates their qualitative understanding: 
\begin{equation}
\label{A4}
\bar{n}(0)\ =\ \frac{\sum_i n(0)}{\sum_i}\ =
\ 1\ +\kappa \rho\ \ ,\ \ \ 
\kappa\ =\ \frac{44}{135}\ \frac{\alpha^2}{m^4}\ \ .
\end{equation}
Here, the sum (integral) $\sum_i$ extends over all directions of propagation 
and all polarizations, $\rho$ and $\alpha$ are the (vacuum) energy 
density and the fine structure constant, respectively.
This sum rule has further been studied very recently in \cite{ditt,gies}
within an elegant formalism which, so far, rigorously only applies to 
electromagnetic background fields, but can be
extended to other situations.

\section{Propagation of light in a constant homogeneous magnetic field}

In this section we want to illustrate the complicated behaviour of 
modified QED vacua by means of the example of a vacuum 
in the presence of a constant 
homogeneous magnetic background field (denoted by ${\bf B}_0$). 
\begin{figure}[htb]
\SetScale{1.0} 
\SetWidth{0.8} 
\SetOffset(176,35) 
\begin{picture}(140,90) 
\Vertex(31.5,15){2} 
\Vertex(108.5,15){2} 
\Vertex(85.5,27){2} 
\Vertex(85.5,3){2} 
\Vertex(54.5,27){2} 
\Vertex(54.5,3){2} 
\Photon(85.5,27)(93.3,56){1}{4.3} 
\Photon(85.5,3)(93.3,-26){1}{4.3} 
\Photon(54.5,27)(46.7,56){1}{4.3} 
\Photon(54.5,3)(46.7,-26){1}{4.3} 
\Line(91.3,54)(95.3,58) 
\Line(95.3,54)(91.3,58) 
\Line(91.3,-28)(95.3,-24) 
\Line(95.3,-28)(91.3,-24) 
\Line(44.7,54)(48.7,58) 
\Line(48.7,54)(44.7,58) 
\Line(44.7,-28)(48.7,-24) 
\Line(48.7,-28)(44.7,-24) 
\CArc(70,-31)(60,50,130) 
\CArc(70,61)(60,230,310) 
\end{picture}
\nopagebreak
\caption{\label{magnpic}
Typical 1-loop diagram contributing to the effective Maxwell action
in the presence of a magnetic background field (indicated
by wavy lines)}
\end{figure}
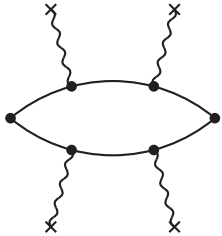
\noindent
For weak magnetic fields ${\bf B}_0$ and
in the low frequency domain $\omega\ll m$, the following
results have been obtained for the refractive index 
at the 1-loop level (cf.\ Fig.\ \ref{magnpic};
$\Theta = \angle ({\bf B}_0,{\bf k})$, ${\bf k}$ is the wave vector
of the test photon) \cite{toll,bial,adle}.
\begin{eqnarray}
\label{B1}
n_\parallel (0) &=&1\ +\frac{\displaystyle 8 \alpha^2}{\displaystyle 45}\ 
\frac{\displaystyle{\bf B}^2_0}{\displaystyle m^4}\ \sin^2\Theta\ \ \  ,
\ \ \ {\bf E}\parallel {\bf k}\times{\bf B}_0\\[3mm]
\label{B2}
n_\perp (0) &=&1\ +\frac{\displaystyle 14 \alpha^2}{\displaystyle 45}\ 
\frac{\displaystyle{\bf B}^2_0}{\displaystyle m^4}\ \sin^2\Theta\ \ ,
\ \ \ {\bf B}\parallel {\bf k}\times{\bf B}_0
\end{eqnarray}
One immediately recognizes that, except for the case of propagation 
parallel to the magnetic field ${\bf B}_0$, the refractive indices for the 
two possible photon polarizations are different.
Consequently, the QED vacuum in the presence of a magnetic (background)
field is a birefringent medium! 
In the case of a constant homogeneous magnetic field, also (1-loop)
results for the refractive index at arbitrary frequencies $\omega$ are 
available \cite{toll,erbe1,erbe2,adle,tsai1,tsai2}. 
Figure \ref{tollpic} shows the behaviour of the two different 
refractive indices over a wide range of 
frequencies (we restrict our consideration to the real part of the 
refractive index). Figure \ref{erberpic} 
displays the polarization averaged refractive
indices related to the phase and group velocities.
No doubt, the graphs look 
quite similar to the graphs
for the refractive index of some, say, condensed matter medium.
It deserves to be emphasized that significant deviations from 
Eqs.\ (\ref{B1}), (\ref{B2}) occur for frequencies $\omega\gg m$
only. From Figs.\ \ref{tollpic}, \ref{erberpic} one recognizes that for 
sufficiently large frequencies $\omega$ the phase and 
group velocities of light
become larger than $c$, but the front velocity of light 
remains equal to $c$ (i.e., $n(\infty)$ is found to be equal to 1;
incidentally, note that with increasing
frequency the validity of QED perturbation theory is subject to
scrutiny).
\begin{center}
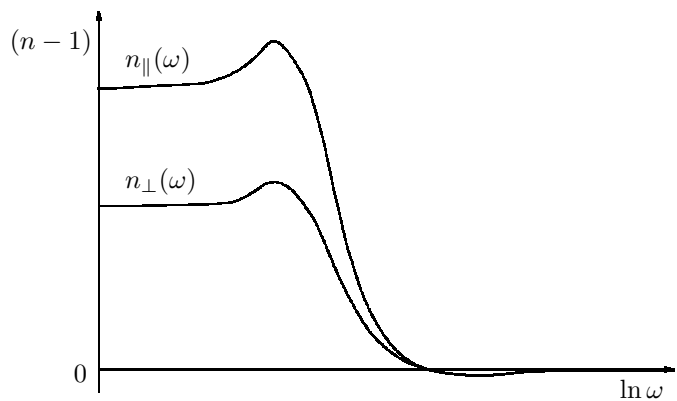
\begin{figure}[htb]
\setlength{\unitlength}{0.00033333in}
\begingroup\makeatletter\ifx\SetFigFont\undefined%
\gdef\SetFigFont#1#2#3#4#5{%
  \reset@font\fontsize{#1}{#2pt}%
  \fontfamily{#3}\fontseries{#4}\fontshape{#5}%
  \selectfont}%
\fi\endgroup%
{\renewcommand{\dashlinestretch}{30}
\begin{picture}(9458,6200)(0,-10)
\drawline(390,129)(390,6069)
\drawline(420.000,5949.000)(390.000,6069.000)(360.000,5949.000)
\drawline(390,3039)
(456.740,3039.000)
(514.263,3039.000)
(563.537,3039.000)
(605.527,3039.000)
(671.528,3039.000)
(720.000,3039.000)
(762.960,3038.911)
(817.496,3038.764)
(872.034,3038.734)
(915.000,3039.000)
(969.653,3040.225)
(1039.011,3042.249)
(1108.364,3044.399)
(1163.000,3046.000)
(1200.013,3046.919)
(1247.000,3048.034)
(1293.987,3049.132)
(1331.000,3050.000)
(1368.233,3050.944)
(1415.496,3052.169)
(1462.761,3053.309)
(1500.000,3054.000)
(1567.500,3053.977)
(1635.000,3054.000)
(1691.194,3055.198)
(1762.511,3057.177)
(1833.823,3059.318)
(1890.000,3061.000)
(1957.504,3062.824)
(2025.000,3065.000)
(2066.443,3067.060)
(2119.030,3069.994)
(2171.602,3073.180)
(2213.000,3076.000)
(2267.475,3079.734)
(2336.571,3084.672)
(2405.632,3091.024)
(2460.000,3099.000)
(2528.269,3119.441)
(2595.000,3144.000)
(2657.679,3172.590)
(2719.000,3204.000)
(2792.682,3251.559)
(2865.000,3301.000)
(2923.384,3344.281)
(2985.000,3384.000)
(3051.548,3403.470)
(3120.000,3414.000)
(3172.969,3409.973)
(3225.000,3399.000)
(3267.591,3380.133)
(3318.439,3352.170)
(3367.567,3321.372)
(3405.000,3294.000)
(3451.901,3242.835)
(3495.000,3189.000)
(3548.190,3122.047)
(3600.000,3054.000)
(3634.406,3005.347)
(3677.190,2943.052)
(3718.879,2879.982)
(3750.000,2829.000)
(3768.337,2794.139)
(3789.430,2750.914)
(3812.228,2702.058)
(3835.680,2650.305)
(3858.734,2598.389)
(3880.340,2549.044)
(3899.445,2505.003)
(3915.000,2469.000)
(3940.406,2407.946)
(3955.743,2370.083)
(3971.811,2330.116)
(3987.840,2290.134)
(4003.061,2252.228)
(4028.000,2191.000)
(4052.215,2133.082)
(4083.171,2059.659)
(4114.542,1986.406)
(4140.000,1929.000)
(4170.531,1864.828)
(4189.779,1825.382)
(4210.255,1783.878)
(4230.886,1742.451)
(4250.602,1703.238)
(4283.000,1640.000)
(4319.869,1570.909)
(4342.949,1528.396)
(4367.466,1483.668)
(4392.175,1439.043)
(4415.830,1396.842)
(4455.000,1329.000)
(4489.605,1271.748)
(4534.324,1199.659)
(4580.630,1128.490)
(4620.000,1074.000)
(4673.112,1014.428)
(4743.626,942.454)
(4816.078,872.503)
(4875.000,819.000)
(4918.261,783.882)
(4974.377,741.180)
(5031.555,699.888)
(5078.000,669.000)
(5126.989,641.409)
(5190.181,608.881)
(5254.283,578.163)
(5306.000,556.000)
(5345.861,541.946)
(5394.961,526.361)
(5450.291,510.019)
(5508.839,493.693)
(5567.594,478.154)
(5623.547,464.178)
(5673.686,452.535)
(5715.000,444.000)
(5757.941,436.969)
(5810.580,429.893)
(5869.702,422.994)
(5932.091,416.497)
(5994.531,410.626)
(6053.806,405.604)
(6106.701,401.654)
(6150.000,399.000)
(6188.663,397.500)
(6235.973,396.439)
(6289.047,395.812)
(6345.004,395.614)
(6400.960,395.839)
(6454.033,396.481)
(6501.340,397.537)
(6540.000,399.000)
(6584.751,402.101)
(6639.394,407.138)
(6700.625,413.503)
(6765.142,420.589)
(6829.642,427.789)
(6890.820,434.495)
(6945.374,440.101)
(6990.000,444.000)
(7059.372,448.239)
(7102.244,450.441)
(7147.459,452.587)
(7192.678,454.598)
(7235.566,456.392)
(7305.000,459.000)
(7341.439,460.054)
(7386.072,461.106)
(7436.166,462.132)
(7488.993,463.107)
(7541.820,464.007)
(7591.917,464.805)
(7636.554,465.478)
(7673.000,466.000)
(7728.578,466.752)
(7787.893,467.502)
(7851.564,468.257)
(7920.206,469.023)
(7994.435,469.804)
(8033.838,470.202)
(8074.868,470.606)
(8117.604,471.017)
(8162.122,471.435)
(8208.498,471.861)
(8256.811,472.296)
(8307.137,472.741)
(8359.554,473.195)
(8414.137,473.660)
(8470.965,474.137)
(8530.114,474.625)
(8591.662,475.127)
(8655.685,475.642)
(8722.261,476.171)
(8791.466,476.716)
(8863.378,477.275)
(8938.073,477.852)
(8976.488,478.146)
(9015.629,478.445)
(9055.504,478.748)
(9096.122,479.055)
(9137.495,479.367)
(9179.631,479.684)
(9222.540,480.006)
(9266.231,480.332)
(9310.715,480.664)
(9356.000,481.000)
\drawline(390,4873)	
(459.777,4874.949)
(519.918,4876.669)
(571.433,4878.193)
(615.334,4879.554)
(684.332,4881.911)
(735.000,4884.000)
(777.707,4886.206)
(829.991,4889.248)
(888.657,4892.857)
(950.515,4896.764)
(1012.371,4900.700)
(1071.032,4904.398)
(1123.306,4907.587)
(1166.000,4910.000)
(1207.982,4912.070)
(1259.408,4914.409)
(1317.130,4916.926)
(1378.000,4919.530)
(1438.870,4922.130)
(1496.592,4924.636)
(1548.018,4926.956)
(1590.000,4929.000)
(1646.198,4932.033)
(1717.522,4936.080)
(1788.835,4940.337)
(1845.000,4944.000)
(1884.690,4946.671)
(1935.053,4950.184)
(1985.389,4954.354)
(2025.000,4959.000)
(2064.196,4966.451)
(2113.491,4977.456)
(2162.541,4989.483)
(2201.000,5000.000)
(2243.905,5014.120)
(2297.754,5033.468)
(2351.225,5053.831)
(2393.000,5071.000)
(2429.150,5088.040)
(2474.370,5110.816)
(2519.155,5134.435)
(2554.000,5154.000)
(2593.760,5179.058)
(2643.282,5212.110)
(2692.164,5246.107)
(2730.000,5274.000)
(2802.929,5335.880)
(2873.000,5401.000)
(2921.253,5457.669)
(2970.000,5514.000)
(3024.810,5567.195)
(3090.000,5611.000)
(3135.641,5613.537)
(3180.000,5604.000)
(3244.717,5555.685)
(3300.000,5499.000)
(3331.915,5460.775)
(3369.874,5410.628)
(3406.646,5359.666)
(3435.000,5319.000)
(3457.510,5284.739)
(3485.230,5240.765)
(3512.335,5196.409)
(3533.000,5161.000)
(3552.034,5125.104)
(3575.230,5079.104)
(3597.811,5032.802)
(3615.000,4996.000)
(3632.534,4955.190)
(3653.899,4903.066)
(3674.565,4850.660)
(3690.000,4809.000)
(3702.551,4769.567)
(3717.348,4719.221)
(3731.721,4668.765)
(3743.000,4629.000)
(3757.948,4576.173)
(3776.706,4509.056)
(3795.362,4441.911)
(3810.000,4389.000)
(3820.050,4352.685)
(3832.781,4306.579)
(3845.371,4260.434)
(3855.000,4224.000)
(3863.989,4188.061)
(3874.764,4143.992)
(3886.684,4094.493)
(3899.106,4042.262)
(3911.391,3989.999)
(3922.895,3940.402)
(3932.979,3896.169)
(3941.000,3860.000)
(3948.751,3822.152)
(3958.150,3774.026)
(3967.474,3725.887)
(3975.000,3688.000)
(3982.216,3653.392)
(3991.168,3611.031)
(4001.297,3563.505)
(4012.043,3513.403)
(4022.845,3463.312)
(4033.146,3415.823)
(4042.384,3373.523)
(4050.000,3339.000)
(4064.773,3273.738)
(4074.013,3233.440)
(4083.798,3190.955)
(4093.607,3148.476)
(4102.923,3108.195)
(4118.000,3043.000)
(4132.699,2979.314)
(4141.787,2939.965)
(4151.380,2898.474)
(4160.988,2856.986)
(4170.121,2817.647)
(4185.000,2754.000)
(4193.693,2716.813)
(4204.263,2671.254)
(4216.136,2620.124)
(4228.740,2566.226)
(4241.501,2512.363)
(4253.847,2461.335)
(4265.204,2415.947)
(4275.000,2379.000)
(4293.856,2315.877)
(4306.053,2277.062)
(4319.104,2236.196)
(4332.267,2195.369)
(4344.799,2156.668)
(4365.000,2094.000)
(4376.667,2056.801)
(4390.788,2011.193)
(4406.614,1960.000)
(4423.391,1906.043)
(4440.370,1852.142)
(4456.799,1801.120)
(4471.926,1755.799)
(4485.000,1719.000)
(4512.796,1648.730)
(4530.656,1605.644)
(4549.837,1560.367)
(4569.352,1515.233)
(4588.210,1472.572)
(4620.000,1404.000)
(4647.400,1351.154)
(4683.621,1284.946)
(4720.782,1219.265)
(4751.000,1168.000)
(4789.549,1106.327)
(4813.764,1068.539)
(4839.632,1028.915)
(4865.921,989.551)
(4891.400,952.546)
(4935.000,894.000)
(4981.533,841.277)
(5042.903,777.120)
(5106.571,715.153)
(5160.000,669.000)
(5204.660,639.228)
(5263.054,605.228)
(5322.420,573.113)
(5370.000,549.000)
(5444.137,516.938)
(5486.218,500.580)
(5520.000,489.000)
(5585.581,472.470)
(5626.272,463.652)
(5669.273,454.965)
(5712.364,446.782)
(5753.328,439.478)
(5820.000,429.000)
(5863.013,423.991)
(5915.704,419.012)
(5974.857,414.228)
(6037.256,409.804)
(6099.686,405.904)
(6158.930,402.693)
(6211.773,400.337)
(6255.000,399.000)
(6294.789,398.690)
(6343.464,399.103)
(6398.058,400.119)
(6455.605,401.616)
(6513.139,403.474)
(6567.694,405.571)
(6616.303,407.787)
(6656.000,410.000)
(6689.959,412.650)
(6731.464,416.598)
(6777.997,421.424)
(6827.041,426.708)
(6876.081,432.028)
(6922.598,436.963)
(6964.077,441.094)
(6998.000,444.000)
(7067.373,448.214)
(7110.247,450.400)
(7155.462,452.535)
(7200.683,454.540)
(7243.571,456.338)
(7313.000,459.000)
(7348.648,460.103)
(7392.310,461.244)
(7441.315,462.390)
(7492.993,463.506)
(7544.671,464.559)
(7593.679,465.515)
(7637.346,466.340)
(7673.000,467.000)
(7718.387,467.729)
(7776.000,468.549)
(7833.613,469.344)
(7879.000,470.000)
(7926.535,470.797)
(7984.752,471.843)
(8050.091,473.043)
(8118.993,474.300)
(8187.895,475.520)
(8253.237,476.607)
(8311.459,477.465)
(8359.000,478.000)
(8427.067,478.466)
(8464.786,478.649)
(8505.451,478.799)
(8549.425,478.915)
(8597.074,478.998)
(8648.762,479.048)
(8704.856,479.065)
(8765.720,479.048)
(8831.719,478.998)
(8903.219,478.915)
(8941.145,478.861)
(8980.584,478.799)
(9021.580,478.728)
(9064.179,478.649)
(9108.427,478.562)
(9154.370,478.466)
(9202.053,478.362)
(9251.522,478.250)
(9302.823,478.129)
(9356.000,478.000)
\drawline(383,485)(9353,485)
\drawline(9233.000,455.000)(9353.000,485.000)(9233.000,515.000)
\put(800,3324){\makebox(0,0)[lb]{\smash{$n_\perp(\omega)$}}}
\put(800,5214){\makebox(0,0)[lb]{\smash{$n_\parallel (\omega)$}}}
\put(0,300){\makebox(0,0)[lb]{\smash{$0$}}}
\put(-1000,5500){\makebox(0,0)[lb]{\smash{$(n-1)$}}}
\put(8520,39){\makebox(0,0)[lb]{\smash{$\ln\omega $}}}
\end{picture}
}
\caption{\label{tollpic}
Qualitative behaviour of the refractive index $n(\omega)$, 
the drawing is adapted from Fig.\ 3.5A of Ref.\
\protect\cite{toll}, p.\ 91}
\end{figure}
\end{center}
\begin{center}
\begin{figure}[htb]
\setlength{\unitlength}{0.00033333in}
\begingroup\makeatletter\ifx\SetFigFont\undefined%
\gdef\SetFigFont#1#2#3#4#5{%
  \reset@font\fontsize{#1}{#2pt}%
  \fontfamily{#3}\fontseries{#4}\fontshape{#5}%
  \selectfont}%
\fi\endgroup%
{\renewcommand{\dashlinestretch}{30}
\begin{picture}(10553,6200)(0,-10)
\drawline(390,99)(390,6039)
\drawline(420.000,5919.000)(390.000,6039.000)(360.000,5919.000)
\drawline(386,4573)
(430.431,4572.520)
(471.747,4572.078)
(545.653,4571.301)
(608.961,4570.655)
(662.913,4570.130)
(708.750,4569.714)
(747.716,4569.394)
(810.000,4569.000)
(854.566,4568.862)
(909.148,4568.797)
(970.407,4568.789)
(1035.004,4568.820)
(1099.600,4568.873)
(1160.858,4568.933)
(1215.437,4568.980)
(1260.000,4569.000)
(1310.505,4569.000)
(1372.363,4569.000)
(1441.789,4569.000)
(1515.000,4569.000)
(1588.211,4569.000)
(1657.637,4569.000)
(1719.495,4569.000)
(1770.000,4569.000)
(1826.182,4569.000)
(1897.500,4569.000)
(1968.818,4569.000)
(2025.000,4569.000)
(2094.401,4569.000)
(2137.282,4569.000)
(2182.500,4569.000)
(2227.718,4569.000)
(2270.599,4569.000)
(2340.000,4569.000)
(2383.079,4569.000)
(2435.841,4569.000)
(2495.057,4569.000)
(2557.500,4569.000)
(2619.943,4569.000)
(2679.159,4569.000)
(2731.921,4569.000)
(2775.000,4569.000)
(2827.873,4568.847)
(2894.993,4568.591)
(2962.116,4568.540)
(3015.000,4569.000)
(3052.159,4570.065)
(3097.646,4571.825)
(3148.685,4574.047)
(3202.500,4576.500)
(3256.315,4578.953)
(3307.354,4581.175)
(3352.841,4582.935)
(3390.000,4584.000)
(3436.267,4584.191)
(3494.993,4583.790)
(3553.721,4583.494)
(3600.000,4584.000)
(3638.158,4585.961)
(3686.530,4589.200)
(3734.888,4592.589)
(3773.000,4595.000)
(3815.956,4596.586)
(3870.492,4598.156)
(3925.032,4600.148)
(3968.000,4603.000)
(4002.838,4607.134)
(4046.811,4613.394)
(4090.629,4620.956)
(4125.000,4629.000)
(4165.128,4643.871)
(4214.485,4665.526)
(4263.099,4688.669)
(4301.000,4708.000)
(4335.808,4728.063)
(4379.084,4754.809)
(4421.818,4782.400)
(4455.000,4805.000)
(4522.905,4859.046)
(4590.000,4914.000)
(4648.834,4960.350)
(4710.000,5004.000)
(4761.653,5028.536)
(4815.000,5049.000)
(4889.284,5067.161)
(4931.276,5074.924)
(4965.000,5079.000)
(5005.080,5079.069)
(5055.341,5076.128)
(5105.432,5070.872)
(5145.000,5064.000)
(5203.472,5045.477)
(5275.069,5017.226)
(5345.381,4986.112)
(5400.000,4959.000)
(5444.955,4931.923)
(5500.031,4895.171)
(5553.842,4856.584)
(5595.000,4824.000)
(5637.231,4783.146)
(5687.891,4728.979)
(5737.106,4673.572)
(5775.000,4629.000)
(5813.110,4580.680)
(5859.949,4518.334)
(5905.563,4455.071)
(5940.000,4404.000)
(5968.317,4355.338)
(6002.171,4292.595)
(6034.940,4229.304)
(6060.000,4179.000)
(6081.796,4132.730)
(6107.735,4075.766)
(6136.310,4011.620)
(6166.015,3943.801)
(6195.343,3875.820)
(6222.788,3811.185)
(6246.842,3753.409)
(6266.000,3706.000)
(6291.282,3638.788)
(6306.364,3597.086)
(6322.034,3553.033)
(6337.512,3508.912)
(6352.019,3467.011)
(6375.000,3399.000)
(6390.731,3350.233)
(6409.558,3290.385)
(6430.364,3223.125)
(6452.033,3152.119)
(6473.447,3081.035)
(6493.491,3013.543)
(6511.047,2953.308)
(6525.000,2904.000)
(6534.525,2868.492)
(6545.789,2824.910)
(6558.167,2775.935)
(6571.039,2724.247)
(6583.780,2672.529)
(6595.770,2623.462)
(6606.384,2579.725)
(6615.000,2544.000)
(6628.661,2486.131)
(6645.060,2415.182)
(6654.002,2376.165)
(6663.287,2335.513)
(6672.802,2293.771)
(6682.432,2251.485)
(6692.065,2209.199)
(6701.586,2167.459)
(6710.882,2126.809)
(6719.839,2087.796)
(6736.280,2016.856)
(6750.000,1959.000)
(6762.358,1907.553)
(6777.447,1844.546)
(6794.474,1773.858)
(6812.650,1699.370)
(6831.182,1624.960)
(6849.278,1554.509)
(6866.148,1491.896)
(6881.000,1441.000)
(6895.788,1394.980)
(6914.636,1338.968)
(6936.419,1276.376)
(6960.014,1210.610)
(6984.294,1145.081)
(7008.135,1083.197)
(7030.412,1028.367)
(7050.000,984.000)
(7082.984,919.179)
(7104.392,879.860)
(7127.558,838.762)
(7151.341,798.007)
(7174.602,759.714)
(7215.000,699.000)
(7246.665,659.599)
(7288.823,611.859)
(7333.069,565.939)
(7371.000,532.000)
(7422.642,497.162)
(7490.116,458.196)
(7559.782,423.132)
(7618.000,400.000)
(7656.060,390.876)
(7702.501,383.322)
(7754.535,377.195)
(7809.372,372.354)
(7864.226,368.656)
(7916.308,365.959)
(7962.828,364.121)
(8001.000,363.000)
(8040.901,362.862)
(8089.632,363.963)
(8144.241,366.043)
(8201.770,368.840)
(8259.266,372.095)
(8313.773,375.547)
(8362.336,378.935)
(8402.000,382.000)
(8464.013,388.558)
(8502.245,393.210)
(8542.537,398.271)
(8582.826,403.361)
(8621.043,408.099)
(8683.000,415.000)
(8726.232,418.707)
(8779.197,422.832)
(8838.655,427.154)
(8901.369,431.454)
(8964.099,435.509)
(9023.606,439.099)
(9076.653,442.003)
(9120.000,444.000)
(9181.180,446.110)
(9251.630,447.945)
(9291.152,448.770)
(9333.975,449.538)
(9380.428,450.254)
(9430.840,450.921)
(9485.538,451.544)
(9544.850,452.127)
(9609.105,452.673)
(9678.630,453.187)
(9715.472,453.433)
(9753.754,453.673)
(9793.518,453.906)
(9834.805,454.134)
(9877.656,454.357)
(9922.111,454.575)
(9968.212,454.789)
(10016.000,455.000)
\drawline(386,4573)
(425.298,4573.000)
(461.840,4573.000)
(527.207,4573.000)
(583.199,4573.000)
(630.916,4573.000)
(671.456,4573.000)
(705.918,4573.000)
\drawline(918.500,4573.000)
(963.718,4573.000)
(1006.599,4573.000)
(1076.000,4573.000)
(1138.793,4573.000)
(1177.589,4573.000)
(1218.500,4573.000)
\drawline(1396.651,4573.000)
(1440.315,4573.000)
(1489.322,4573.000)
(1541.000,4573.000)
(1592.678,4573.000)
(1641.685,4573.000)
(1685.349,4573.000)
\drawline(1856.000,4573.000)
(1894.758,4573.000)
(1931.513,4573.000)
(1991.000,4573.000)
(2050.487,4573.000)
(2087.242,4573.000)
(2126.000,4573.000)
\drawline(2328.492,4572.655)
(2396.000,4573.000)
(2436.586,4574.817)
(2488.049,4577.825)
(2539.487,4581.170)
(2580.000,4584.000)
(2628.736,4587.573)
\drawline(2837.519,4608.831)
(2883.657,4617.246)
(2929.717,4626.039)
(2966.000,4633.000)
(3021.385,4643.308)
(3091.672,4656.480)
(3161.874,4670.162)
\drawline(3416.000,4738.000)
(3459.745,4755.322)
(3514.426,4778.975)
(3568.644,4803.641)
(3611.000,4824.000)
(3647.948,4843.723)
(3694.209,4869.789)
\drawline(3911.000,5008.000)
(3978.811,5067.700)
(4046.000,5128.000)
(4104.432,5174.819)
(4166.000,5218.000)
(4225.145,5243.016)
(4286.000,5263.000)
\drawline(4505.346,5287.305)
(4554.735,5285.644)
(4594.000,5281.000)
(4657.548,5263.206)
(4695.432,5250.039)
(4734.791,5234.941)
(4773.677,5218.601)
\drawline(4999.000,5026.000)
(5043.770,4960.298)
(5069.837,4918.895)
(5096.659,4874.878)
(5122.960,4830.563)
(5147.468,4788.268)
(5186.000,4719.000)
\drawline(5269.000,4539.000)
(5291.002,4488.146)
(5318.460,4423.409)
(5345.438,4358.467)
(5366.000,4307.000)
(5387.573,4247.793)
(5400.457,4211.068)
\drawline(5460.000,4037.000)
(5479.123,3979.995)
(5503.146,3907.549)
(5527.096,3835.078)
(5546.000,3778.000)
(5569.249,3708.640)
(5583.658,3665.798)
\drawline(5651.000,3463.000)
(5661.758,3428.941)
(5674.782,3387.184)
(5689.277,3340.284)
(5704.449,3290.792)
(5719.502,3241.265)
(5733.641,3194.254)
\drawline(5783.279,3016.547)
(5793.965,2975.616)
(5804.592,2934.671)
(5814.649,2895.836)
(5831.000,2833.000)
(5847.743,2770.268)
(5858.143,2731.524)
\drawline(5906.000,2548.000)
(5922.540,2474.592)
(5932.349,2429.156)
(5942.524,2381.210)
(5952.571,2333.238)
(5961.995,2287.723)
(5977.000,2214.000)
\drawline(6008.854,2043.474)
(6017.852,1994.490)
(6026.473,1948.054)
(6034.321,1906.709)
(6041.000,1873.000)
(6052.838,1817.583)
(6068.392,1747.361)
\drawline(6128.122,1492.294)
(6146.101,1419.878)
(6161.000,1363.000)
(6173.359,1319.775)
(6189.635,1265.106)
(6206.593,1210.634)
(6221.000,1168.000)
\drawline(6311.000,958.000)
(6329.679,924.205)
(6354.631,882.141)
(6401.000,808.000)
(6431.997,760.432)
(6471.935,700.596)
(6513.905,641.963)
\drawline(6707.297,471.996)
(6754.404,439.490)
(6800.224,410.011)
(6842.506,385.276)
(6879.000,367.000)
(6935.330,348.258)
(7007.371,331.412)
\drawline(7243.737,306.104)
(7286.184,306.306)
(7328.608,306.970)
(7368.835,307.923)
(7434.000,310.000)
(7503.528,313.615)
(7546.428,316.392)
\drawline(7749.000,333.000)
(7782.986,337.166)
(7824.493,342.961)
(7871.013,349.851)
(7920.038,357.301)
(7969.057,364.776)
(8015.563,371.741)
\drawline(8197.872,392.148)
(8240.944,395.773)
(8284.022,399.301)
(8324.878,402.602)
(8391.000,408.000)
(8460.377,413.990)
(8503.241,417.708)
\drawline(8706.000,434.000)
(8745.494,436.256)
(8793.874,438.693)
(8848.180,441.200)
(8905.451,443.670)
(8962.729,445.994)
(9017.053,448.062)
\drawline(9233.881,453.704)
(9272.884,454.334)
(9313.514,454.939)
(9355.230,455.517)
(9397.489,456.065)
(9439.748,456.579)
(9481.465,457.058)
\drawline(9690.000,459.000)
(9746.698,459.253)
(9782.171,459.316)
(9823.900,459.337)
(9873.017,459.316)
(9930.652,459.253)
(9997.936,459.148)
%
\drawline(383,459)(10305,459)
\drawline(10185.012,428.952)(10305.000,459.000)(10184.988,488.952)
\put(6390,4074){\makebox(0,0)[lb]{\smash{$n(\omega)$}}}
\put(9615,39){\makebox(0,0)[lb]{\smash{$\ln\omega $}}}
\put(4320,3159){\makebox(0,0)[lb]{\smash{$n_{\rm gr}(\omega)$}}}
\put(0,300){\makebox(0,0)[lb]{\smash{$0$}}}
\put(-1000,5500){\makebox(0,0)[lb]{\smash{$(n-1)$}}}
\end{picture}
}
\caption{\label{erberpic}
Qualitative behaviour of the polarization averaged refractive indices
$n(\omega)$, $n_{\rm gr}(\omega)$ (note that the arbitrary units 
applied to the axes are different here from those used in 
Fig.\ \protect\ref{tollpic});
the drawing is adapted from Fig.\ 2
of Ref.\ \protect\cite{erbe1}, p.\ 711; 
also see \protect\cite{erbe2}, Sec.\ 5.A}
\end{figure}
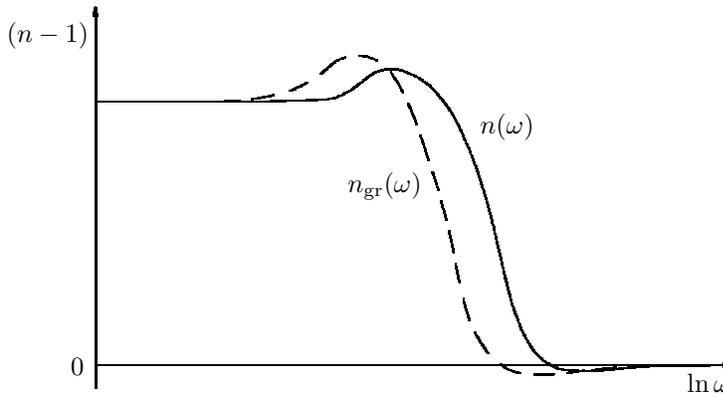
\end{center}

\section{\protect\label{CASISEC}The Casimir vacuum}

\begin{figure}[htb]
\unitlength1.mm
\begin{picture}(140,40)
\put(115,12){$x_3$}
\put(82,33){$L$}
\put(60,20){\vector(1,0){60}}
\put(70,30){\vector(1,0){30}}
\put(100,30){\vector(-1,0){30}}
\linethickness{0.7mm}
\put(70,5){\line(0,1){30}}
\put(100,5){\line(0,1){30}}
\end{picture}
\caption{\label{casifig}
The Casimir vacuum: Two uncharged parallel conducting plates 
enforcing at their surfaces the standard boundary conditions 
${\bf E}_\parallel = 0 = {\bf B}_\perp$
on photon vacuum fluctuations}
\end{figure}
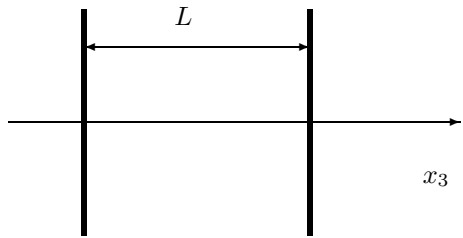

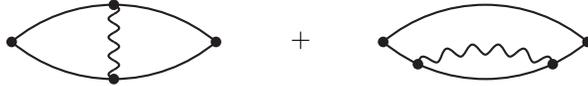
\begin{figure}[htb]
\SetScale{1.0} 
\SetWidth{0.8} 
\SetOffset(135,5) 
\unitlength1pt 
\begin{picture}(140,40) 
\Vertex(1.5,15){2} 
\Vertex(78.5,15){2} 
\Vertex(40,29){2} 
\Vertex(40,1){2} 
\Photon(40,29)(40,1){2}{4} 
\CArc(40,-31)(60,50,130) 
\CArc(40,61)(60,230,310) 
\put(242,18){$+$} 
\Vertex(141.5,15){2} 
\Vertex(218.5,15){2} 
\Vertex(205.4,6.6){2} 
\Vertex(154.6,6.6){2} 
\PhotonArc(180,-47.8)(60,65,115){2}{5.3} 
\CArc(180,-31)(60,50,130) 
\CArc(180,61)(60,230,310) 
\end{picture}
\nopagebreak
\caption{\label{casidiag} 2-loop diagrams contributing 
in a nontrivial way to the effective Maxwell action in a Casimir vacuum}
\end{figure}

The Casimir effect \cite{casi,milo2,most} is a macroscopic quantum 
effect which is considered to be a manifestation of the existence
of (photon) vacuum fluctuations. In its best known form, 
it consists in the mutual
attraction of two parallel uncharged conducting plates (mirrors)
in vacuo separated by a distance $L$ (cf.\ Fig.\ \ref{casifig}) 
according to the law
\begin{eqnarray}
p&=& -\ \frac{\pi^2}{240}\ \frac{1}{L^4}\ =\
-\ \frac{\partial E_{\rm vac}}{\partial L}\ \ ,
\ \ \ E_{\rm vac}\ =\ -\ \frac{\pi^2}{720}\ \frac{1}{L^3}\ \ ,
\end{eqnarray}
where $p$ is the force per unit area (Casimir pressure) and 
$E_{\rm vac}$ is the vacuum energy per unit area of the mirrors
(the vacuum energy density $\rho$ is given by $E_{\rm vac}/L$).
\noindent
The effect has recently been verified quantitatively in
the laboratory \cite{lamo,mohi}. It is the result of the 
boundary conditions enforced on the photon vacuum fluctuations 
at the mirror surfaces\footnote{As the Casimir effect is an 
infrared (i.e., long distance)
phenomenon its physics is fairly independent of the fact that 
in reality the boundary conditions cease to apply beyond the 
cut-off frequency of the mirrors. This comment also applies 
to the discussion of the propagation of light in a Casimir vacuum
including the problem of the front velocity of light in it.}.
The change in the vacuum structure effected by the parallel mirrors
which manifests itself in terms of the Casimir effect also affects
the propagation of light in the Casimir vacuum, as can be recognized from 
Eq.\ (\ref{A4}) (by the term `Casimir vacuum' we denote
the vacuum between the mirrors). In order to study this problem
one has to calculate the effective Maxwell action for the given 
situation. This has been done in \cite{scha,bart1} in an 
approximation which considers plates distances $L$ much larger
than the electron Compton wavelength $m^{-1}$ and low 
frequency test photons ($\omega\ll m$). The result can
reliably be calculated within QED perturbation theory by taking
into account the diagrams shown in Fig.\ \ref{casidiag}.
\noindent
For the effective Maxwell action one finds the following result (between the mirrors).
\begin{eqnarray}
\Gamma_{\rm eff}[{\bf E},{\bf B}]\ =\ \frac{1}{2}\int{\rm d}^4 x
\left(\epsilon_{ij}E^i E^j\right.  &-&\left. \mu^{-1}_{\ \ ij}B^i B^j\right)\\
\left\{
\begin{array}{r}
\epsilon_{11}\ =\ \epsilon_{22}\\
\mu_{33}
\end{array}
\right\}& =& 1\ 
\left\{
\begin{array}{c}
+\\
-
\end{array}
\right\}\ 
\frac{\pi^2}{180}\ \frac{\alpha^2}{(mL)^4}\ \left[g(x_3) - 
\frac{11}{45}\right]\nonumber\\
\left\{
\begin{array}{r}
\epsilon_{33}\\
\mu_{11}\ =\ \mu_{22}
\end{array}
\right\}&=&1\ 
\left\{
\begin{array}{c}
+\\
-
\end{array}
\right\}\ 
\frac{\pi^2}{180}\ \frac{\alpha^2}{(mL)^4}\ \left[g(x_3) + 
\frac{11}{45}\right]\nonumber
\end{eqnarray}
Here, $g(x_3)$ is some explicitly known function which 
the diagonal elements of the permittivity and permeability tensors
(all non-diagonal elements are zero)
depend on but not the refractive index itself 
calculated in WKB approximation.
For the low frequency refractive index one finds
\begin{eqnarray}
n_\parallel(0) &=& 1
\end{eqnarray}
for propagation of light parallel to the mirrors (as is expected from 
the residual Lorentz invariance with respect to boosts parallel
to the mirrors) and
\begin{eqnarray}
n_\perp (0) &=&1\ -\ \frac{11\pi^2}{(90)^2}\ 
\frac{\alpha^2}{(mL)^4}
\end{eqnarray}
for propagation perpendicular to the mirrors (in the space between them).
In the approximation used the propagation of light is non-dispersive,
consequently the phase and group velocities agree and are given by
$c/n_\perp (0) > c$ (incidentally, analogous results have been
found in the case of gravitational background fields). 
The change in these velocities related to the 
presence of the parallel mirrors is interesting as a matter of principle,
but it is unmeasurably small in practice. Moreover,
it cannot be detected by test waves whose Fourier components 
respect $\omega\ll m$ \cite{milo1,benm,bart3}. 

In qualitative respect,
of most interest appears to be the question what can be said
about the front velocity of light $c/n_\perp (\infty)$ in the Casimir vacuum.
So far, no calculation of the refractive index $n(\omega)$ for
arbitrary frequencies $\omega$ has been performed (an attempt
for a complete 2-loop calculation made in \cite{lato} has turned
out to be incorrect). Quite generally, the calculation of
the infinite frequency limit of the refractive index is 
a nonperturbative task the technical tools we are presently 
lacking for\footnote{The point is that for $\omega\ll m$ one can 
rely on a derivative expansion of the effective Maxwell action.
For the Casimir vacuum, e.g., the 2-loop contribution to $n(0)$
yields the dominant term if $mL\gg 1$ applies. 
However, for $\omega\rightarrow\infty$
($\omega \gg m$) the derivative expansion is no longer appropriate.
Then, loops of arbitrary order are contributing in principle. }. 
However, in the present case there is an alternative to the explicit 
calculation. This alternative approach is based on structural 
arguments which are linked to basic physical principles \cite{bart3}.
The only nonperturbative relation at hand is the standard
Kramers-Kronig dispersion relation for the refractive index
which embodies the principle of local causality, i.e., the 
principle that no effect can precede its cause. The standard 
Kramer-Kronig relation reads
\begin{equation}
\mbox{Re}\ n(\omega)\ =\ \mbox{Re}\ n(\infty)\ +\ \frac{2}{\pi}
\int\limits^\infty_0{\rm d}\omega^\prime\ 
\frac{\omega^\prime\ \mbox{Im}\ n(\omega^\prime)}{\omega^{\prime 2}
- \omega^2}
\end{equation}
and can immediately be specialized to the following form we are
mainly interested in (for a more detailed discussion see \cite{bart3}).
\begin{equation}
\label{C6}
n(\infty)\ =\ n(0)\ -\ \frac{2}{\pi}
\int\limits^\infty_0\frac{{\rm d}\omega^\prime}{\omega^\prime}\ 
\mbox{Im}\ n(\omega^\prime)
\end{equation}
If one assumes that the Casimir vacuum behaves like a 
passive medium ($\mbox{Im}\ n_\perp(\omega) \ge 0$) from 
Eq.\ (\ref{C6}) one finds the inequality
\begin{equation}
\label{A7}
1\ >\ n_\perp (0)\ \ge\ n_\perp (\infty)\ \ . 
\end{equation}
Eq.\ (\ref{A7}) provides us with an upper bound on $n_\perp (\infty)$
irrespective of any nonperturbative contribution. Consequently, we 
may conclude (we denote this conclusion by the term `alternative A'): 
\begin{itemize}
\item[(A.)]
The front velocity of light
$c/n_\perp (\infty)$ in a Casimir
vacuum is larger than the front velocity of light $c$ in the unbounded 
space vacuum .
\end{itemize}
As unconventional as this conclusion may appear to
be, it does not contradict the special theory of relativity. In this
context, it is important to realize that the parallel mirrors (i.e., the 
Casimir vacuum) define a distinguished reference system. The equivalence
of all inertial systems special relativity is based on no longer 
applies (this point, however, is not exclusively 
related to the Casimir vacuum, or other modified vacua, and
can also be found in discussions of classical media, cf.\ \cite{bolo},
e.g.). Furthermore, the above conclusion does not entail any violation 
of the principle of local causality\footnote{This, however, is 
controversially being discussed. For a recent paper in this 
respect see \cite{dolg3}.}. 
A well defined (but modified) light cone continues to exist. 
And finally, it is the etalon phenomenon
itself which the space-time structure is defined by which is modified
(``light does never move faster than light'').
In other words, the light cone related to $c$ is no longer of any
physical significance and, therefore, cannot be used to construct
thought experiments exhibiting some causality violation.

The above conclusion (alternative A) 
with respect to the front velocity of light
$c/n_\perp (\infty)$ in a Casimir vacuum relies on Eq.\ (\ref{A7})
which has been derived on the basis of a few assumptions. 
The alternative A can 
only be evaded if these assumptions were not valid. Consequently,
mainly there are two (further) {\it logical} alternatives to it \cite{bart3}: 
\begin{itemize}
\item[B.]
The standard Kramers-Kronig
relation fails to be valid in the present context.

\item[C.]
$\mbox{Im}\ n_\perp(\omega) < 0$, at least for some range
of frequencies $\omega$. In this case the Casimir vacuum would 
behave like an active medium amplifying the test wave. This 
seems to be in conflict with the principle of energy conservation
unless a physical explanation removing this concern could be found
(presently, for it even not a guess exists).
\end{itemize}
If one wished to avoid the alternative A one would have to face the
logical alternatives B or C. It should, however, be emphasized that the latter
alternatives would entail fairly unconventional physics.

\section{Conclusions}

QED vacua under the influence of external conditions
are complicated dispersive media.
The appearance of (phase and group) velocities larger than $c$
is a common phenomenon in these vacua.
The front velocity of light $c/n(\infty)$ in such 
vacua cannot be calculated by means of presently available
theoretical tools -- this is a truly nonperturbative task.
Regarding the special case of the Casimir vacuum, it can be said that the 
low frequency refractive index $n_\perp (0)$ for propagation perpendicular
to two parallel mirrors (in the space between them) 
has reliably been calculated within QED 
perturbation theory and found to be smaller than 1.
This result can be used indirectly to infer information 
about the related front velocity by
relying on the standard Kramers-Kronig relation which 
embodies the principle of local causality 
(i.e., the fact that there can be no effect
preceding its cause). If one then conventionally assumes
that the Casimir vacuum behaves like a 
passive medium ($\mbox{Im}\ n_\perp(\omega) \ge 0$)
one is led to conclude that the front velocity of light
$c/n_\perp (\infty)$ in a Casimir
vacuum is larger than the front velocity of light $c$ in the unbounded 
space vacuum\footnote{One should bear in mind, however, that 
this conclusion depends on a few 
assumptions which might fail to apply, although this would
entail fairly unconventional and, therefore, equally 
interesting physics as pointed out at the end of 
Sec.\ \ref{CASISEC}.}. It should finally be emphasized that
this conclusion does not involve any serious conceptual dangers
and does not, in particular,  contradict the special theory of relativity.

\end{document}